# Switching of the Chiral Magnetic Domains in the Hybrid Multiferroic (ND$_4$)$_2$[FeCl$_5$(D$_2$O)]

J. Alberto Rodríguez-Velamazán,[1*] Oscar Fabelo,[1*] Javier Campo,[2] Juan Rodríguez-Carvajal,[1] Navid Qureshi[1] and Laurent C. Chapon[1,3]

[1] *Institut Laue-Langevin, 71 Avenue des Martyrs, CS 20156, 38042 Grenoble Cedex 9, France.*

[2] *Instituto de Ciencia de Materiales de Aragón, CSIC-Universidad de Zaragoza, C/ Pedro Cerbuna 12, E-50009, Zaragoza, Spain.*

[3] *Diamond Light Source Ltd, Harwell Science & Innovation Campus, Didcot OX11 0DE, Oxon, England*

*Corresponding Authors: velamazan@ill.eu, fabelo@ill.eu*

**Abstract**

Neutron spherical polarimetry, which is directly sensitive to the absolute magnetic configuration and domain population, has been used in this work to unambiguously prove the multiferroicity of (ND$_4$)$_2$[FeCl$_5$(D$_2$O)]. We demonstrate that the application of an electric field upon cooling results in the stabilization of a single-cycloidal magnetic domain below 6.9 K, while poling in the opposite electric field direction produces the full population of the domain with opposite magnetic chirality. We prove the complete switchability of the magnetic domains at low temperature by the applied electric field, which constitutes a direct proof of the strong magnetoelectric coupling. Additionally, we refine the magnetic structure of the ordered ground state, determining the underlying magnetic space group consistent with the direction of the ferroelectric polarization, and we provide evidence of a collinear amplitude-modulated state with magnetic moments along the *a*-axis in the temperature region between 6.9 and 7.2 K.

Unlike "type I" multiferroics, where electric and magnetic orders coexist but are weakly coupled, the magneto-electric (ME) coupling is very strong in "type II" (or "improper") multiferroics, where the magnetic ordering breaks spatial inversion symmetry and induces electric polarization [1,2,3]. This direct coupling leads to remarkable possibilities of manipulation of the electric order by magnetic fields and of the magnetic order by electric fields, which are at the origin of the significant attention devoted to this class of materials. The number of such spin-driven multiferroics is, however, still scarce, and mainly limited to transition metal oxides like RMnO$_3$ (R= rare earth) [4,5,6] MnWO$_4$ [7,8], CoCr$_2$O$_4$ [9], CuFeO$_2$ [10] or RbFe(MoO$_4$)$_2$ [11]. The difficult control of the delicate balance between exchange couplings in these systems impedes a rational design of these interesting materials. This fact highlights the interest for innovative approaches for obtaining magnetically driven multiferroic compounds. Important efforts have been made in this sense following metal-organic routes, which allow the combination of building blocks with the desired properties in a bottom-up approach.[12,13,14,15] However, the cross coupling between magnetic and electric orders is absent or very weak in this type of systems.[16,17] As an alternative, hybrid approaches inspired by the architectures of the successful organic-inorganic photovoltaic materials have been proposed.[18] For this research, microscopic techniques that give direct proof of the multiferroicity and give insight on the underlying magneto-electric coupling mechanism are crucial, since a good understanding of these mechanisms may help to design new improper multiferroic materials with improved functional properties.

The family $A_2[FeX_5(H_2O)]$,[19] where A stands for an alkali metal or ammonium ion and X for a halide ion, represents a new route to obtain materials with strong ME coupling. The occurrence of ME effects in the alkali-based compounds $K_2[FeCl_5(H_2O)]$, $Rb_2[FeCl_5(H_2O)]$ and $Cs_2[FeCl_5(H_2O)]$ has been described in a thorough study of their macroscopic physical properties,[20] and understood from the analysis of their crystal and magnetic structures[21, 22]. While the previous compounds are linear magneto-electric materials, i.e. where the electric polarization can be induced by applying a magnetic field, another member of the family, $(NH_4)_2[FeCl_5(H_2O)]$[18], has been proven to be a veritable spin-driven multiferroic in the ground state, i.e. the appearance of an electric polarization is spontaneous and directly related to magnetic ordering. We recently determined the mechanisms of multiferroicity in the deuterated form of this material in different regions of its rich magnetic field-temperature (*B-T*) phase diagram from a detailed determination of its crystal and magnetic structures by neutron diffraction [23, 24]. In absence of magnetic field, this compounds becomes magnetically ordered at 7.2 K ($T_N$) and ferroelectric below 6.9 K ($T_{FE}$). The magnetic structure in the ground state is cycloidal spiral propagating along the *c*-axis and with magnetic moments mainly contained in the *ac*-plane. A ferroelectric polarization, primarily directed along the *a*-axis, develops in this phase, which is compatible with the spin current mechanism [23]. In a scenario reminiscent of that observed in canonical compounds like $TbMnO_3$ or $MnWO_4$,[4-8] the symmetry of the magnetic structure in the ground state is consistent with two successive transitions (as indicated by two anomalies observed in the low temperature region of the heat capacity), suggesting that two order parameters condense in turn at $T_{FE}$ and $T_N$. This would correspond to a transition on warming from a cycloidal (T < $T_{FE}$) to a collinear antiferromagnetic structure ($T_{FE}$ < T < $T_N$), the latter being compatible with the absence of electric polarization.

In the present work, we prove the possibility of controlling the "chiral" magnetic domains of this material by an applied electric field (E-field). We use neutron spherical polarimetry to determine the absolute magnetic configuration and domain population of this system under different external E-fields. The cycloidal magnetic structure of the ground state implies two "chiral" domains with opposite rotation of the cycloids (Fig.1), which directly correspond to electric domains with opposite polarities. Therefore, by applying an E-field, we can manipulate the population of the "chiral" magnetic domains, and we show that the system is completely switchable, which represents a direct evidence of its multiferroic character. In addition, we use neutron spherical polarimetry data to refine the previously proposed magnetic structure of the ground state, obtaining a more accurate magnetic structure model. Finally, we confirm the existence of a collinear sinusoidal magnetic phase with magnetic moments along the *a*-axis in the temperature region between 6.9 and 7.2 K.[25]

Spherical neutron polarimetry[26] is an ideally suited technique to study complex magnetic structures. This is a zero magnetic field technique which allows to access the direction and phase of the magnetic scattering, unreachable by unpolarized methods. The experimental setup allows setting the polarization of the incident neutron beam in an arbitrary direction and analyzing the polarization of the scattered beam in another spatial direction.[27] In the present case, with a magnetic propagation vector $\mathbf{k} = (0, 0, k_z)$ [$k_z = 0.23$], the magnetic and nuclear Bragg peaks never superimpose, and therefore the polarization of the scattered beam for each Bragg peak can be calculated, for any incident polarization, merely from the magnetic structure factor $\mathbf{M}(\mathbf{Q})$ (a complex vector).[28,29] The most convenient setup for these calculations is to use the Blume reference frame,[28] with the *X*-axis parallel to the scattering vector $\mathbf{Q}$, the *Z*-axis perpendicular to the scattering plane (vertical in our case), and the *Y*-axis completing the right-hand set. Magnetic neutron diffraction is only sensitive to the component of $\mathbf{M}(\mathbf{Q})$ perpendicular to $\mathbf{Q}$, $\mathbf{M}_\perp(\mathbf{Q})$, the so-called magnetic interaction vector, therefore lying in the *YZ* plane: $\mathbf{M}_\perp(\mathbf{Q}) = (0, M_{\perp y}, M_{\perp z})$. The scattering of a fully polarized incident neutron beam can be expressed by the polarization matrix *P*, with matrix elements $P_{ij}$ *(i,j = X,Y,Z)*, representing the polarization of the scattered beam in the direction *j*, for an incident beam polarized in the direction *i* (see Supporting Information for details). In the present case, for a pure magnetic Bragg peak probed at $\mathbf{Q}$ (for simplicity, we will drop the argument $\mathbf{Q}$ in the expressions and consider a perfectly polarised neutron beam):

$$P_{mag} = \begin{pmatrix} -1 & 0 & 0 \\ \frac{2\text{Im}(M_{\perp y}M_{\perp z}^*)}{M_\perp^2} & -\frac{M_{\perp z}^2 - M_{\perp y}^2}{M_\perp^2} & \frac{2\text{Re}(M_{\perp y}M_{\perp z}^*)}{M_\perp^2} \\ \frac{2\text{Im}(M_{\perp y}M_{\perp z}^*)}{M_\perp^2} & \frac{2\text{Re}(M_{\perp y}M_{\perp z}^*)}{M_\perp^2} & \frac{M_{\perp z}^2 - M_{\perp y}^2}{M_\perp^2} \end{pmatrix}$$

The off-diagonal components, $P_{yx}$, $P_{zx}$, are the so-called 'chiral' terms, and can only be non-zero for non-collinear structures. Their value not only depends on the magnetic structure, but also on the populations of the possible magnetic domains, hence the particular sensitivity of spherical neutron polarimetry to the domain structure. In particular, in the case of cycloidal structures as the one of the ground state of the present compound, the rotation of the cycloid is opposite for each "chiral" domain, and thus the signs of the 'chiral' terms are reversed for the same magnetic reflection probed at **Q**. For a single domain, the 'chiral' terms will have a finite value, with the sign indicating the handedness (clockwise or counter-clockwise) of the "chiral" magnetic domain. Finally, the chirality of a single domain is only observable, when the projection of the cycloidal envelope along Q is non-zero. The sense of rotation of the cycloid is coupled with the direction of the ferroelectric polarization through the spin current mechanism[23], and therefore the application of an external E-field can modify the populations of the described magnetic domains. If both domains are equally populated, the 'chiral' terms average out and $P_{yx} = P_{zx} = 0$ (as would be the case for a collinear structure), but if the domain populations are unbalanced, which can be realized by cooling in an applied E-field in the direction of the ferroelecetric polarization, finite values are measured, which allow the determination of the relative fraction of cycloidal domains.

For collinear arrangements, like the proposed at $T_{FE} < T < T_N$, off-diagonal terms are zero, but the diagonal terms contain information about the orientation of the magnetic moments that is often undistinguishable by unpolarized neutron diffraction techniques. In the general case, the final values of both the diagonal and off-diagonal terms for each magnetic reflection further depend on the details of the magnetic structure and the exact orientation of the moments with respect to **Q**. In helical structures, the diagonal $P_{yy}$ and $P_{zz}$ components contain information about the ellipticity (for example, for a reflection with the magnetic moment nearly perpendicular to **Q**, these terms are close to zero when the envelope is circular), while the off-diagonal terms $P_{yz}$ and $P_{zy}$ are sensitive to the inclination of the plane of rotation of the magnetic moments. Therefore, the measurement of polarization matrices in a series of magnetic reflections allows for a particularly precise refinement of the magnetic structure (especially if combined with a set of magnetic Bragg peaks intensities obtained by unpolarized neutron diffraction), together with the determination of the relative fraction of spiral domains.

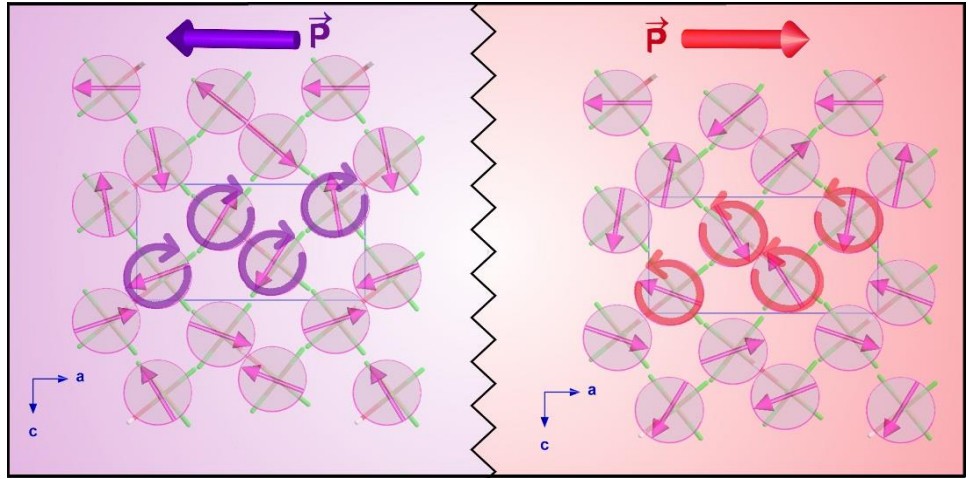

FIG. 1. Schematic representation of the magnetic structure of the two magnetic domains, with the direction of the resulting electric polarization.

A large prism-shaped single crystal of $(ND_4)_2[FeCl_5(D_2O)]$ ($P112_1/a$, with $a \approx 3.50$ Å, $b \approx 9.96$ Å and $c \approx 6.90$ Å and $\gamma \approx 90.1°$ in the low temperature phase) of dimensions ca. 2 x 3 x 4 mm along the crystallographic $a$-, $b$- and $c$- directions was obtained by the seeded growth technique as described in our previous work [23]. Spherical neutron polarimetry experiments were conducted on the hot neutron diffractometer D3 at the Institut Laue-Langevin (ILL, Grenoble, France), equipped with the CryoPAD device,[27] operating at $\lambda = 0.84$ Å. The sample was mounted onto a sample stick allowing applying up to 10 kV high voltage by a potential difference between two parallel horizontal aluminum plates. The crystal was fixed to the lower plate by silver epoxy with the $a^*$-axis in the vertical direction (see Figure S1), and the upper plate positioned at ca. 1 mm from the sample surface (total distance between electrodes, 3.15 mm, sample thickness, ca. 2 mm). The aluminum sample chamber was indium-sealed and evacuated, and installed into the CryoPAD device at D3 instrument. Measurements were carried out after either zero-field cooling or electric-field cooling through the multiferroic transition. The crystal alignment with an accuracy better than 0.5 degrees was obtained by a previous orientation of the crystal using the neutron Laue diffractometer Orient Express[30]. We have probed several magnetic reflections, (0 1 $k_z$), (0 1 -$k_z$) and (0 0 1-$k_z$), selected considering their magnetic signal and geometrical conditions to have the propagation vector in the scattering plane and a significant component of the magnetic moment perpendicular to **Q** (see inset of Fig.2 top). Unpolarized neutron diffraction data were collected at the hot-neutrons four-circle diffractometer D9 at ILL ($\lambda = 0.832$ Å). The crystal was placed into a closed-cycle cryostat and 127 independent magnetic reflections were recorded at 2 K (together with a large set of nuclear reflections that was used to accurately set the scale factor). We used the Mag2Pol program[31] for analyzing and fitting the spherical neutron polarimetry data jointly with unpolarized neutron diffraction intensities. The calculated polarization matrix components were corrected for the imperfect incident neutron beam polarization and the efficiency of the neutron spin-filter in the scattered beam. The spin-filter efficiency was monitored by measuring the polarization term $P_{zz}$ on the purely nuclear Bragg peak (0 2 0).

The spherical neutron polarimetry results in the $T_{FE} < T < T_N$ phase are displayed in Fig. 2a – right (measurements at T = 7 K after zero-E-field cooling). As expected, the polarization matrix elements of the (0 1 $k_z$) reflection have non-zero values only for the diagonal terms, $P_{ii}$, consistent with the magnetic structure being collinear, and thereby sinusoidal, since it remains incommensurate [23]. The external electric field does not affect the neutron polarization matrix in this non-ferroelectric phase (the only domains in this phase are the indistinguishable anti-phase domains, corresponding with a π/2 shift on the sinusoidal structure, and not coupled to the electric polarization). The symmetry lowering from the paramagnetic phase to the ground-state cycloidal phase indicates that two order parameters condense successively at $T_{FE}$ and $T_N$. Depending on what is the first one that condenses, two sinusoidal structures are possible in the $T_{FE} < T < T_N$ phase, either with moments along *a* or along *c*. In our configuration, the $P_{yy}$ term is particularly sensitive to the magnetic moments being along *a* or along *c* directions. Our results (Fig. 2 – right) univocally discard the second option, and therefore are consistent with a collinear sinusoidal magnetic phase with magnetic moments directed along *a*.[25] Fig. 2b displays the temperature dependence of selected terms of the polarization matrix, showing how the value of the off-diagonal $P_{yx}$ term becomes zero when entering in the 6.9 K < T < 7.2 K phase, as corresponds to its collinear character, while the diagonal terms evolve to absolute values close to one, with signs consistent with a collinear sinusoidal magnetic structure with the magnetic moments directed along *a*.

We examined in detail the magnetic structure and the domain population in the cycloidal phase at T < $T_{FE}$ by measuring polarization matrices for a series of magnetic reflections in both zero-E-field cooled and E-field cooled conditions (positive and negative fields along the *a\** direction). The results corresponding to measurements of the (0 1 0.23) magnetic reflection at 4 K are shown in Fig. 2a – left, while Fig. 2b shows the temperature dependence of selected polarization matrix terms. Further results for other magnetic reflections, entirely consistent with the ones in Fig. 2, are provided in the Supplemental Material. The results of the full polarization matrix measurement for the single (0 1 0.23) magnetic reflection immediately evidence:

- (i) The finite values of the 'chiral' components, $P_{yx}$ (and $P_{zx}$), confirming the non-collinear character of the magnetic structure in this phase and the E-field induced unbalance of domain populations.

- (ii) The change of sign of these terms when the sample is cooled in either negative or positive E-field (while the rest of the terms remain the same), which demonstrates the reversal of the domain populations by a reversal of the E-field.

A joint fit of data of three full polarization matrices and the intensities of 127 independent magnetic reflections obtained by unpolarized neutron diffraction (see figures S2 and S3 in Supplemental Material) yields a refined model of the magnetic structure in the ground state consisting on a cycloid with an ellipticity of 0.97(2), and an angle between the rotation plane of the magnetic moments and the *ac*-plane of 2.0(2)° (the full set of parameters of the refinement is given in table 1). This model slightly improves the fit with respect to the previously reported in ref. [23]. The four Fe(III) atoms in the primitive unit-cell, labeled as Fe(1), Fe(2), Fe(3) and Fe(4), with crystallographic coordinates (0.388, 0.249, 0.313), (0.119, 0.751, 0.813), (0.619, 0.751, 0.687) and (0.881, 0.249, 0.187) respectively, were grouped in two orbits where the magnetic moments were constrained to be equal and the envelopes of the spin cycloid to be circular, in order to avoid over parametrization of the fit. In the present case, the last constraint is lifted, allowing an elliptical envelope. Concerning the symmetry of the system, once it becomes ferroelectric below 6.9 K, the non-polar $P112_1/a$ space group is not strictly correct, although the structural changes responsible of this type of electric polarization are usually too subtle to be observable. However, the magnetic symmetry should account for the polar character of the system. The magnetic space group should be a polar subgroup of the parent space group $P112_1/a1'$. The highest polar symmetry subgroup is $P112_11'(00\gamma)000s$, but this magnetic space group belongs to the 2⁠1′ point group, which only allows ferroelectric polarization along the *c*-axis and consequently is not compatible with the spontaneous electric polarization observed from the pyroelectric measurements.[18] Therefore, the only magnetic super-space group deriving from the parent group compatible with the observed polarization is $P11'(\alpha\beta\gamma)0s$, which allows ferroelectric polarization in whatever direction, and should correspond to the symmetry of our system. In this way, the breaking of symmetry from the parent space group explains the physical properties of the system. The fitting of the domain population gives values of 0.03/0.97(2) for negative E-field cooling and 0.97/0.03(3) for positive E-field cooling, that is, a nearly full population of one or the other domain depending on the sign of the applied E-field. A small magnetic domain unbalance of 0.39/0.61(2) is observed when the sample is cooled without applied E-field, with values of $P_{yx}$ and $P_{zx}$ different from zero.

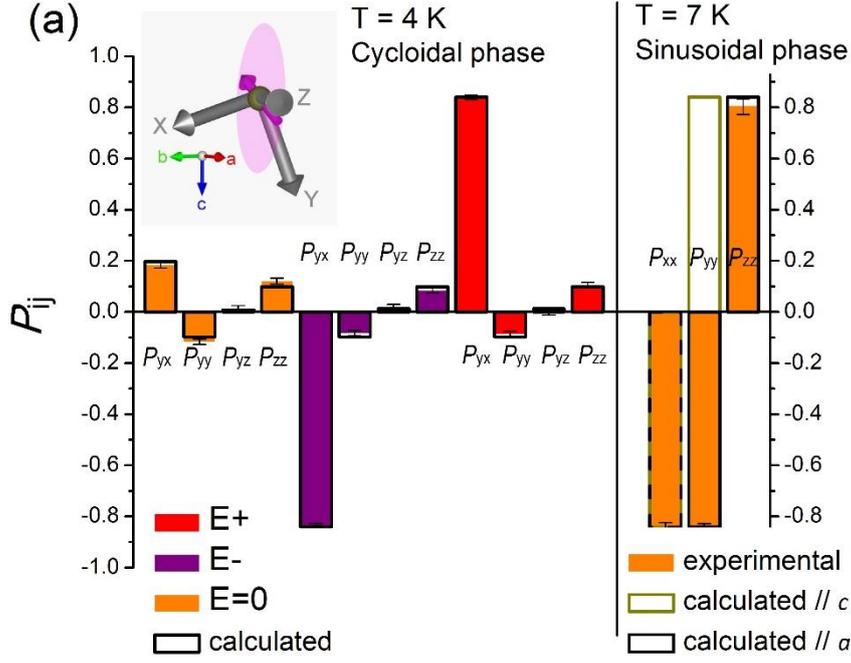

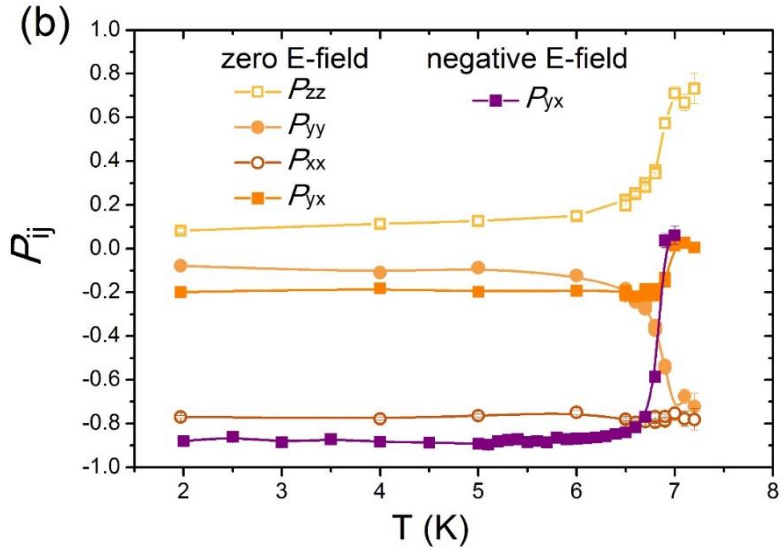

FIG. 2. (a) Observed (solid color columns and error bars) and calculated (black rectangles) neutron spherical polarization matrix elements, $P_{ij}$, for the (0 1 0.23) magnetic reflection of a $(ND_4)_2[FeCl_5(D_2O)]$ crystal oriented with $a^*$ parallel to the Z-axis. Left: Results corresponding to the cycloidal phase (T = 4 K) for the sample cooled in zero electric field (orange) and under a negative (purple) and positive (red) electric field of 25 kV cm$^{-1}$ applied along $a^*$. The calculated values are the results of the joint fit of neutron spherical polarimetry and unpolarized neutron diffraction intensities. Right: Experimental results (orange bars) corresponding to the sinusoidal phase (T = 7 K) for the sample cooled at zero-field, compared with the calculated values for a sinusoidal magnetic structure with moments parallel to $c$ and $a$ directions (green and black rectangles, respectively). Inset: Scheme of the experimental geometry showing the crystal (red/green/blue arrows) and Blume (gray arrows) reference frames, together with the cycloid rotation plane with one particular magnetic moment (both in magenta). (b) Temperature dependence of observed neutron

spherical polarization matrix elements, $P_{ij}$, for the (0 1 0.23) magnetic reflection in zero-E-field cooling and negative E-field cooling conditions.

Once established the coupling between the magnetic domain population and the E-field applied on cooling, we demonstrate how we can switch the magnetic domains by E-field reversal at constant temperature, which constitutes an unambiguous an direct proof of the multiferroicity on this material. Starting from a negative E-field-cooled state, we have followed one of the chiral terms, $P_{yx}$, of the polarization matrix of the (0 1 0.23) magnetic reflection -- directly related with the magnetic domain population -- as a function of a variable E-field (Fig. 3). The magnetic domains are strongly pinned at the lowest temperatures, with the polarization matrix term $P_{yx}$, changing by only ca. 25% for applied positive E-fields up to 25 kV cm$^{-1}$ (corresponding to a domain unbalance of ca. 0.10/0.90). Therefore, we performed the switching measurements at 6.5 K, sufficiently close to the ferroelectric transition. We can clearly observe a complete hysteresis loop, with the population of the magnetic domains fully and reversibly switched by the applied E-field, that is, we have direct control over the magnetic domain population by E-field.

| $R_x$ | $R_y$ | $R_z$ |
|---|---|---|
| -3.96(6) | 0.14(8) | 0 |
| $I_x$ | $I_y$ | $I_z$ |
| -0.01(1) | -0.36(8) | -3.83(6) |

TABLE S1. Refined Fourier components of the magnetic moment, $S_1=1/2(\mathbf{R}+i\mathbf{I})$ and $S_2=1/2(\mathbf{R}+i\mathbf{I})\exp(-2\pi i\phi)$ for the two independent Fe atoms ($R_i$ and $I_i$ are the components along the crystallographic axes) from the joint fit of neutron spherical polarimetry and unpolarized neutron diffraction intensities. Agreement factors: $R_F$ = 12.52 % for the integrated intensities (slightly better than in ref. [23]) and $\chi^2$ = 7.20 for the polarization matrix elements. Both independent magnetic atoms are constrained to have equal values of the components of the magnetic moment, with a refined phase difference of $\phi$=0.495(5) between both.

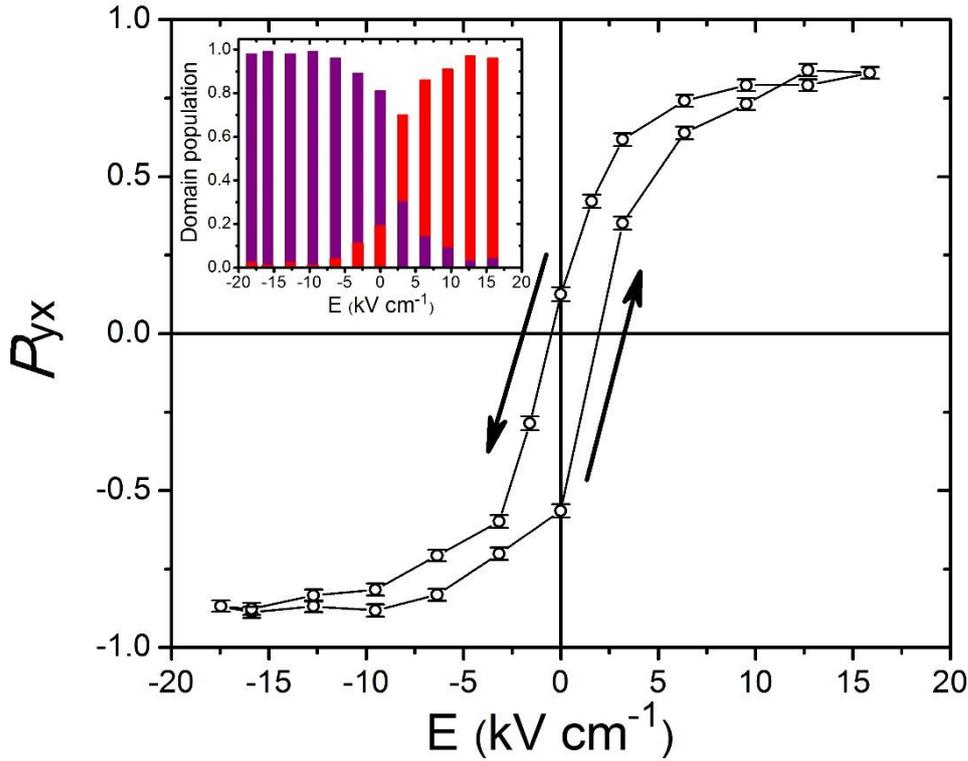

FIG. 3. Hysteresis loop measured on the off-diagonal neutron polarization matrix element $P_{yx}$ of the (0 1 0.23) reflection of a $(ND_4)_2[FeCl_5(D_2O)]$ crystal. The sample was initially cooled down to 4 K under a negative E-field of 17.5 kV cm$^{-1}$ and then warmed to 6.5 K, temperature that was kept constant for the subsequent measurements as a function of a variable E-field. Inset: Domain populations (purple and red bars refer to negative and positive chiral magnetic domains, respectively) corresponding to the increasing E-field branch of the hysteresis loop.

In summary, by means of spherical neutron polarimetry experiments, combined with unpolarized neutron diffraction data, we have determined in detail the magnetic structures of $(ND_4)_2[FeCl_5(D_2O)]$ hybrid compound in both its cycloidal (T < 6.9 K) and sinusoidal (6.9 K < T < 7.2 K) phases. In the temperature region between 6.9 and 7.2 K we have determined a collinear sinusoidal magnetic structure with magnetic moments along the *a*-axis, consistent with the symmetry lowering scheme. At temperatures below 6.9 K, we have obtained an improved model for the magnetic structure of the ground state, determining the underlying magnetic space group, which allows explaining the ferroelectric polarization from symmetry considerations. We are able to stabilize a nearly full population of either chiral magnetic domain by cooling in an external electric field. Furthermore, we can tune the domain populations by varying the electric field, reaching a full population reversal by switching the field in a complete hysteresis loop, thus providing a direct proof at the microscopic level of the multiferroicity of $(ND_4)_2[FeCl_5(D_2O)]$, which represents a promising example of the hybrid molecular/inorganic approach to materials with strong magneto-electric coupling.

Partial funding for this work is provided through the grant MAT2015-68200-C2-2-P. We are grateful to Institut Laue-Langevin for the neutron beam-time allocated in instruments D3 and D9 through project "5-51-503" (doi:10.5291/ILL-DATA.5-51-503).

# Switching of the Chiral Magnetic Domains in the Hybrid Multiferroic $(ND_4)_2[FeCl_5(D_2O)]$

**J. Alberto Rodríguez-Velamazán,[1*] Oscar Fabelo,[1*] Javier Campo,[2] Juan Rodríguez-Carvajal,[1] and Laurent C. Chapon[1, 3]**

[1] *Institut Laue-Langevin, 71 Avenue des Martyrs, CS 20156, 38042 Grenoble Cedex 9, France.*

[2] *Instituto de Ciencia de Materiales de Aragón, CSIC-Universidad de Zaragoza, C/ Pedro Cerbuna 12, E-50009, Zaragoza, Spain.*

[3] *Diamond Light Source Ltd, Harwell Sci & Innovat Campus, Didcot OX11 0DE, Oxon, England*

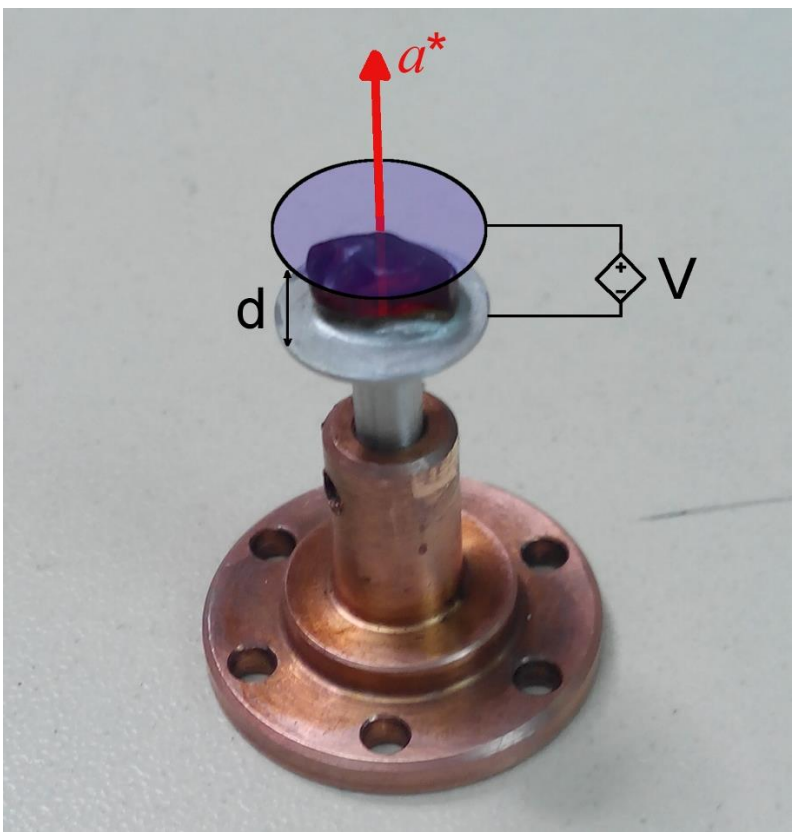

FIG. S1. Single crystal of $(ND_4)_2[FeCl_5(D_2O)]$ used in the spherical neutron polarimetry experiments, mounted in the sample support designed for the application of electric field by a potential difference between two parallel horizontal aluminum plates. The crystal was fixed to the lower plate by silver epoxy with the $a^*$-axis in the vertical direction, and the upper plate positioned at ca. 1 mm from the sample surface (total distance between electrodes, d = 3.15 mm, sample thickness, ca. 2 mm).

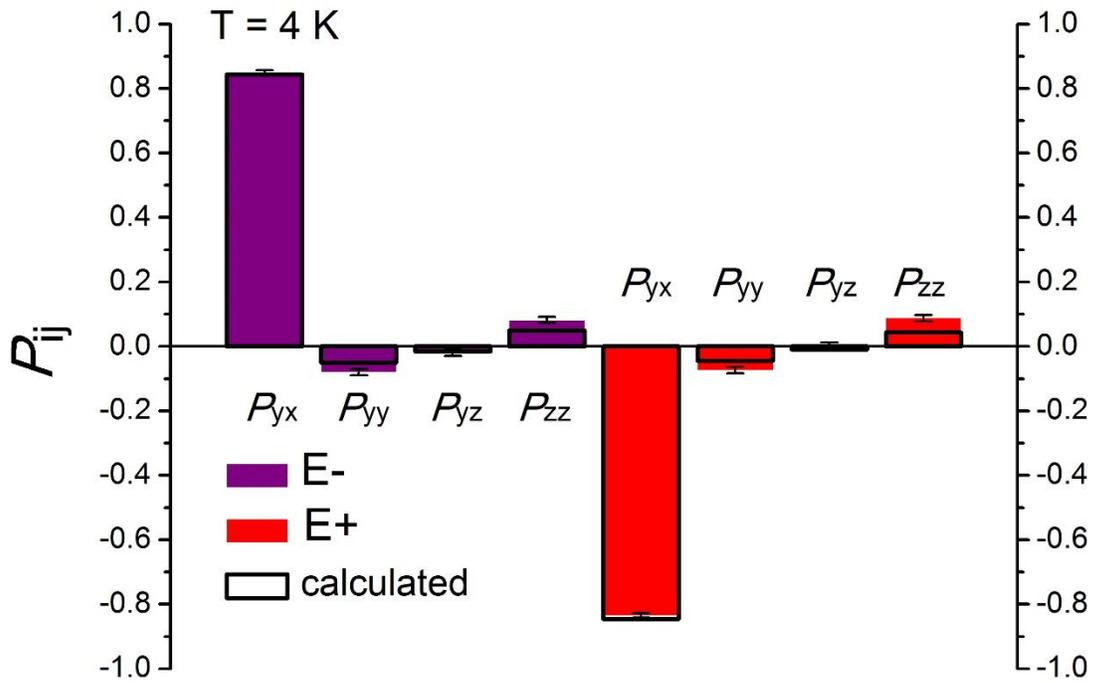

FIG. S2. Observed (solid color columns and error bars) and calculated (black rectangles) neutron spherical polarization matrix elements, $P_{ij}$, for the (0 1 -0.23) magnetic reflection of a $(ND_4)_2[FeCl_5(D_2O)]$ crystal oriented with $a^*$ parallel to the z-axis. Results corresponding to the cycloidal phase (T = 4 K) for the sample cooled under negative (purple) and positive (red) electric field of 25 kV cm$^{-1}$ applied along $a^*$. The calculated values are the results of the joint fit of neutron spherical polarimetry and unpolarized neutron diffraction intensities.

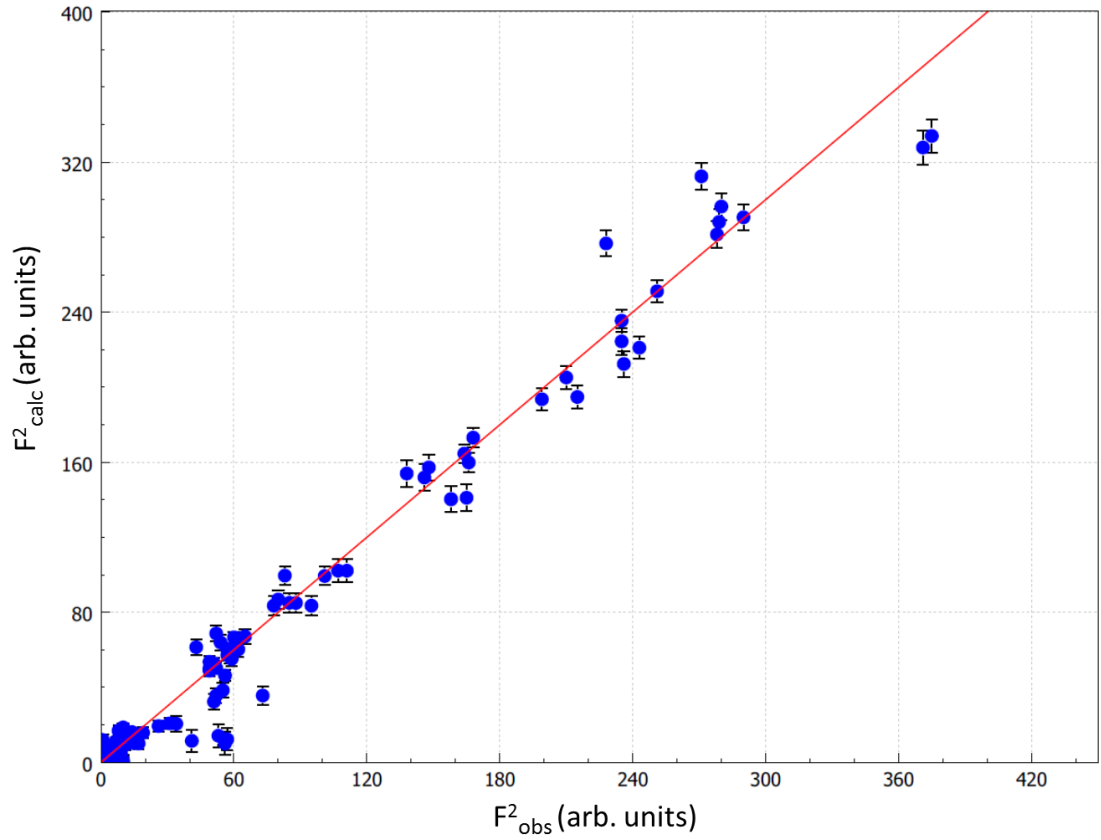

FIG. S3. Results of the joint fit of neutron spherical polarimetry and unpolarized neutron diffraction intensities: Plot of the observed *vs* calculated squared magnetic interaction vectors (here $F^2=|M_\perp|^2$) for unpolarized neutron diffraction (the experimental data are represented as blue circles and the ideal case $F^2_{cal} = F^2_{obs}$, is represented as a solid red line).

**Summary of equations governing the spherical neutron polarimetry technique**

The Blume-Maleyev equations describe the scattering of polarised neutrons. We adopt here a simplified crystallographic notation in order to make clear the important terms. We suppose that we are in the more general case of a crystal with a magnetic structure of arbitrary complexity that has some magneto-elastic coupling and there is a nuclear contribution to the magnetic reflections. We write the Blume-Maleyev equations for a reflection $\mathbf{h}=\mathbf{H}+\mathbf{k}$, where $\mathbf{H}$ is a vector of the average crystallographic reciprocal lattice and the propagation vector $\mathbf{k}$ index the magnetic satellite. We shall call $I_\mathbf{h}$ the intensity (scattering cross-section) of the diffracted beam and, if the incident and scattered polarisations are noted as $\mathbf{P}_i$ and $\mathbf{P}_f$, respectively, the equations are:

Cross-section (Equation 1):

$$I_\mathbf{h} = N_\mathbf{h} N_\mathbf{h}^* + \mathbf{M}_{\perp\mathbf{h}} \mathbf{M}_{\perp\mathbf{h}}^* + \left(N_\mathbf{h} \mathbf{M}_{\perp\mathbf{h}}^* + N_\mathbf{h}^* \mathbf{M}_{\perp\mathbf{h}}\right) \mathbf{P}_i + i\left(\mathbf{M}_{\perp\mathbf{h}}^* \times \mathbf{M}_{\perp\mathbf{h}}\right) \mathbf{P}_i$$

Final polarisation (Equation 2):

$$\mathbf{P}_f I_\mathbf{h} = N_\mathbf{h} N_\mathbf{h}^* \mathbf{P}_i - \left(\mathbf{M}_{\perp\mathbf{h}} \mathbf{M}_{\perp\mathbf{h}}^*\right) \mathbf{P}_i + \left(\mathbf{P}_i\ \mathbf{M}_{\perp\mathbf{h}}^*\right) \mathbf{M}_{\perp\mathbf{h}} + \left(\mathbf{P}_i\ \mathbf{M}_{\perp\mathbf{h}}\right) \mathbf{M}_{\perp\mathbf{h}}^* -$$

$$- i\left(N_\mathbf{h} \mathbf{M}_{\perp\mathbf{h}}^* - N_\mathbf{h}^* \mathbf{M}_{\perp\mathbf{h}}\right) \times \mathbf{P}_i +$$

$$+ N_\mathbf{h} \mathbf{M}_{\perp\mathbf{h}}^* + N_\mathbf{h}^* \mathbf{M}_{\perp\mathbf{h}} - i\left(\mathbf{M}_{\perp\mathbf{h}}^* \times \mathbf{M}_{\perp\mathbf{h}}\right)$$

Where the nuclear structure factor of reflection $\mathbf{h}$ is $N_\mathbf{h}$ and $\mathbf{M}_{\perp\mathbf{h}}$ is the magnetic interaction vector: perpendicular component of the magnetic structure factor to the scattering vector.

The four terms in the scattering cross section are called, respectively: nuclear contribution, magnetic contribution, nuclear-magnetic interference term and chiral term. For a non-polarised beam $\mathbf{P}_i = 0$ only the first two terms contribute to the diffraction pattern. For a pure magnetic reflection only the second and fourth terms are different from zero.

These equations are written in absolute form. They are independent from the particular frame to describe the vectorial quantities. Notice that in the expression of the scattered polarisation the last line regroups the terms that are independent from the incident polarisation. These are the terms that are exploited to obtain polarised beams from an initial non-polarised beam.

We shall consider in the following that we use an arbitrary Cartesian system for referring the component of the vectorial quantities. Let us concentrate in the expression of the scattered polarisation that we will write in a tensorial form. We shall drop the $\mathbf{h}$ and $\perp$ indices to simplify the notation so that the expression of the final polarisation is:

$$\mathbf{P}_f I = NN^* \mathbf{P}_i - (\mathbf{M}\,\mathbf{M}^*)\,\mathbf{P}_i + (\mathbf{P}_i\,\mathbf{M}^*)\mathbf{M} + (\mathbf{P}_i\,\mathbf{M})\mathbf{M}^* - i(N\mathbf{M}^* - N^*\mathbf{M}) \times \mathbf{P}_i +$$
$$+ N\mathbf{M}^* + N^*\mathbf{M} - i(\mathbf{M}^* \times \mathbf{M})$$

The first line can be written in a matrix form and the second is just an added vector. Dividing by the cross-section we can write:
$$\mathbf{P}_f = \vec{P}\mathbf{P}_i + \mathbf{P}_c$$

with
$$\mathbf{P}_c = \frac{N\mathbf{M}^* + N^*\mathbf{M} - i(\mathbf{M}^* \times \mathbf{M})}{I} = \frac{\mathbf{W}_R - \mathbf{T}}{I}$$

being this term independent of the initial polarisation (except for the dependence of $I$) and it is usually called *created* polarization. Let us call the complex vector $\mathbf{W}= 2\,N\mathbf{M}^*= \mathbf{W}_R+i\,\mathbf{W}_I$ the *nuclear-magnetic interference vector*. We can see that $2\mathbf{W}_R=\mathbf{W}+\mathbf{W}^*=2(N\mathbf{M}^*+ N^*\mathbf{M})$ is a real vector and $2\mathbf{W}_I i=\mathbf{W}-\mathbf{W}^*= 2(N\mathbf{M}^*- N^*\mathbf{M})$ is a pure imaginary vector. The vector $\mathbf{M}^*\times \mathbf{M}$ is purely imaginary, so that $\mathbf{T} = i(\mathbf{M}^*\times \mathbf{M})$ is a real vector that we shall call hereafter the *chiral vector*.

In the Blume Cartesian reference system $x$ is along the scattering vector, $z$ is perpendicular to the scattering plane pointing up in the instrument and $y$ completes the right handed system. In this system the intensity reduces to:

$$I = NN^* + \mathbf{M}\mathbf{M}^* + \mathbf{W}_R \mathbf{P}_i + \mathbf{T}\,\mathbf{P}_i = I_N + I_M + (\mathbf{W}_R + \mathbf{T})\mathbf{P}_i$$

$$I = I_N + I_M + T_x P_{ix} + W_{Ry} P_{iy} + W_{Rz} P_{iz}$$

And the matrix equation for polarization is:

$$\mathbf{P}_f = \frac{1}{I}\begin{pmatrix} I_N - I_M & -W_{Iz} & W_{Iy} \\ W_{Iz} & I_N + I_M^y - I_M^z & M_{mix} \\ -W_{Iy} & M_{mix} & I_N - I_M^y + I_M^z \end{pmatrix}\begin{pmatrix} P_{ix} \\ P_{iy} \\ P_{iz} \end{pmatrix} + \frac{1}{I}\begin{pmatrix} -T_x \\ W_{Ry} \\ W_{Rz} \end{pmatrix}$$

where we have put $I_M = \mathbf{M}.\mathbf{M}^* = M_x M_x^* + M_y M_y^* + M_z M_z^* = I_M^x + I_M^y + I_M^z = I_M^y + I_M^z$; the term along $x$ disappears by definition of the reference frame. And we have called $M_{mix} = M_y^* M_z + M_y M_z^* = 2\,\text{Re}(M_y M_z^*)$. The chiral vector is
$$\mathbf{T} = i(\mathbf{M}^* \times \mathbf{M}) = (T_x, 0, 0) = i(M_y M_z^* - M_y^* M_z, 0, 0) = -2(\text{Im}(M_y M_z^*), 0, 0)$$
In practice the incident polarisation is put along $x$, $\mathbf{P}_i=(1,0,0)$, $y$, $\mathbf{P}_i=(0,1,0)$ and $z$, $\mathbf{P}_i=(0,1,0)$, then we measure the three components of the scattered polarisation for each case, so that 9 numbers are obtained for a single reflection. Let us call $I^\alpha$ the scattered intensity when the incident polarisation is along $\alpha$ ($\alpha = x, y, z$). The nine numbers written in the form of a matrix are

$$P = \begin{pmatrix} \dfrac{I_N - I_M - T_x}{I^x} & \dfrac{W_{Iz} + W_{Ry}}{I^x} & \dfrac{W_{Rz} - W_{Iy}}{I^x} \\ \dfrac{-W_{Iz} - T_x}{I^y} & \dfrac{I_N + I_M^y - I_M^z + W_{Ry}}{I^y} & \dfrac{M_{mix} + W_{Rz}}{I^y} \\ \dfrac{W_{Iy} - T_x}{I^z} & \dfrac{M_{mix} + W_{Ry}}{I^z} & \dfrac{I_N - I_M^y + I_M^z + W_{Rz}}{I^z} \end{pmatrix}$$

In which each row corresponds to the final polarization when the initial polarization is along $x$, $y$ and $z$. For the case of a pure magnetic reflection $\mathbf{W}=0$, $I_N=0$, $I^x = I_M + T_x$, $I^y = I^z = I_M$, so the matrix reduces to:

$$P_{mag} = \begin{pmatrix} -1 & 0 & 0 \\ \dfrac{-T_x}{I_M} & \dfrac{I_M^y - I_M^z}{I_M} & \dfrac{M_{mix}}{I_M} \\ \dfrac{-T_x}{I_M} & \dfrac{M_{mix}}{I_M} & \dfrac{I_M^z - I_M^y}{I_M} \end{pmatrix}$$

In terms of the magnetic interaction vector components (we restore here the $\perp$ symbol) $\mathbf{M}_\perp = (0, M_{\perp y}, M_{\perp z})$

$$P_{mag} = \begin{pmatrix} -1 & 0 & 0 \\ \dfrac{2\,\mathrm{Im}(M_{\perp y} M_{\perp z}^*)}{\mathbf{M}_\perp \mathbf{M}_\perp^*} & \dfrac{M_{\perp y}^2 - M_{\perp z}^2}{\mathbf{M}_\perp \mathbf{M}_\perp^*} & \dfrac{2\,\mathrm{Re}(M_{\perp y} M_{\perp z}^*)}{\mathbf{M}_\perp \mathbf{M}_\perp^*} \\ \dfrac{2\,\mathrm{Im}(M_{\perp y} M_{\perp z}^*)}{\mathbf{M}_\perp \mathbf{M}_\perp^*} & \dfrac{2\,\mathrm{Re}(M_{\perp y} M_{\perp z}^*)}{\mathbf{M}_\perp \mathbf{M}_\perp^*} & \dfrac{M_{\perp z}^2 - M_{\perp y}^2}{\mathbf{M}_\perp \mathbf{M}_\perp^*} \end{pmatrix}$$

Or in terms of the real an imaginary components of the magnetic interaction vector

$$\mathbf{M}_\perp = \mathbf{A} + i\mathbf{B} = (0, A_y, A_z) + i(0, B_y, B_z)$$

$$P_{mag} = \begin{pmatrix} -1 & 0 & 0 \\ \dfrac{2(A_y B_z - A_z B_y)}{\mathbf{M}_\perp \mathbf{M}_\perp^*} & \dfrac{A_y^2 + B_y^2 - A_z^2 - B_z^2}{\mathbf{M}_\perp \mathbf{M}_\perp^*} & \dfrac{2(A_y A_z + B_y B_z)}{\mathbf{M}_\perp \mathbf{M}_\perp^*} \\ \dfrac{2(A_y B_z - A_z B_y)}{\mathbf{M}_\perp \mathbf{M}_\perp^*} & \dfrac{2(A_y A_z + B_y B_z)}{\mathbf{M}_\perp \mathbf{M}_\perp^*} & -\dfrac{A_y^2 + B_y^2 - A_z^2 - B_z^2}{\mathbf{M}_\perp \mathbf{M}_\perp^*} \end{pmatrix}$$